\def\apjl{ApJ}%
\def\apjs{ApJS}%
\def\ao{Appl.~Opt.}%
\def\aap{A\&A}%
\documentclass[traditabstract]{aa} 
\usepackage{longtable}
\usepackage{graphicx}
\usepackage{latexsym}
\usepackage{amssymb}
\usepackage{lscape}
\usepackage{txfonts}
\title{Refurbished GOLDMine}

\author{G. Gavazzi \inst{1}
\and P. Franzetti \inst{2}
\and A. Boselli \inst{3}
}
\authorrunning{G. Gavazzi et al.}
\titlerunning{Refurbished GOLDMine}
\institute{Universit\`a degli Studi di Milano-Bicocca, Piazza della Scienza 3, 20126 Milano, Italy\\
\email {gavazzi@mib.infn.it}
\and
INAF - IASF, via Bassini 15, 20133, Milano, Italy\\
\email {paolo@lambrate.inaf.it}
\and
Laboratoire d'Astrophysique de Marseille, UMR6110 CNRS, 38 rue F. Joliot-Curie, F-13388 Marseille France\\
 \email {Alessandro.Boselli@lam.fr}
}

\begin{document}

 
  \abstract
 {We present the refurbished  GOLDMine galaxy site (http://goldmine.mib.infn.it) 
superseding the one operating since 2002 (Gavazzi et al. 2003). 
Data for 9704 galaxies selected
with $r\leq$17.7 mag from the SDSS, belonging to the Cancer cluster, Local supercluster, Coma
supercluster, Hercules, A2197 and A2199 clusters are included.  
}   

   \maketitle
%

\section{GOLDMine}

Galaxies belonging to the various regions of GOLDMine satisfy the conditions
r$\leq$17.7 mag and  (see Fig. 1):\\
Cancer cluster       0$<cz<$9000 $\rm km~s^{-1}$\\
Local supercluster   $cz<$3000 $\rm km ~s^{-1}$\\
Coma supercluster     4000$<cz<$9500 $\rm km ~s^{-1}$\\
A2147+2151 clusters   8000$<cz<$15000 $\rm km ~s^{-1}$\\
A2197+2199 clusters         7000$<cz<$11000 $\rm km ~s^{-1}$

\noindent The new GOLDMine site enables to visualize (and retrieve in tabular form) multifrequency data (from the UV to the centimetric radio,
continuum and line, whose list can be selected with the "available data" button) by navigating (continuum panning and zooming -in and -out) the whole 
northern sky covered by SDSS (DR10) (Ahn et al. 2014) that is set as color- background picture. Alternatively one can choose to pan
other all-sky panoramic (X-ray, UV, optical, FIR and radio) data-sets.
At any time the coordinates of the center of the visible frame are given (in degrees or in hh,mm,ss) and the width 
of the field of view (FOV) is indicated.

Targets can be selected by cluster, by name, by coordinates (both in degrees and in hh,mm,ss),
individually (marked with a red circle) or by lists. For all targets a fast link to the SDSS (DR10) navigator and to the NED database are offered.
By switching-in the "show targets in FOV" option, all targets visible in the FOV belonging to the GOLDMine database are 
highlighted (white circles) and the corresponding data can be retrieved with "get data from FOV".

All FITS images that were available in the original GOLDMine site are still available for download (images panel). 
These include all $u,g,r,i,z$ images from the SDSS, some Johnson B,V,J,H,K frames 
and some spectra taken in the drift-scan mode. 
Many new H$\alpha$ (NET and OFF frames) (Gavazzi et al. 2012, 2013 and Gavazzi et al. 
in preparation) are available.

Comments and criticism are warmly encouraged (e-mail to gavazzi@mib.infn.it). If you make use of GOLDMine, please
continue to cite Gavazzi et al. (2003). 

\begin{figure}[]
\centering
\includegraphics[width=9cm]{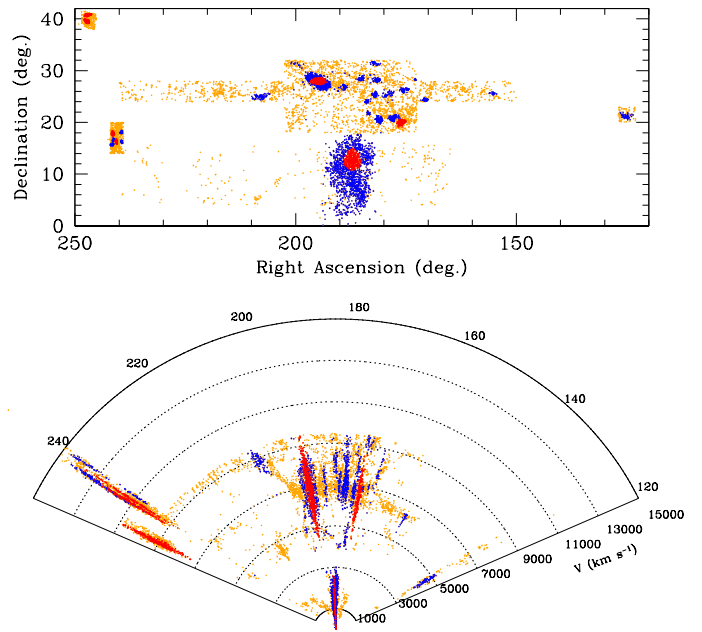}
 \caption{Distribution in celestial coordinates (top panel) and wedge diagram (bottom panel) of 9704 galaxies in the 
 new GOLDMine database. Objects are color-coded (yellow - blue - orange) with increasing local galaxy density.  
 }
 \label{campione} 
\end{figure}

\end{document}